\documentclass[twocolumn,amsmath,amssymb,nofootinbib,superscriptaddress,longbibliography,floatfix]{revtex4-1}

\usepackage{graphicx}
\usepackage{amsmath}
\usepackage{amssymb}
\usepackage{xcolor}
\usepackage{hyperref}
\hypersetup{pdfstartview={FitH},pdfpagemode={UseNone}, breaklinks=true, colorlinks=true, linkcolor=blue, citecolor=blue, urlcolor=blue, bookmarksopen=true, pdfnewwindow=true}
\usepackage[all]{hypcap}
\usepackage[english]{babel}
\usepackage{physics}
\usepackage{braket}

\begin{document}

\preprint{APS/123-QED}

\title{Topological corner states in metasurfaces based on split-ring resonators with dipole-quadrupole couplings}

\author{Alina D. Rozenblit}
\email{alina.rozenblit@metalab.ifmo.ru}
\affiliation{School of Physics and Engineering, ITMO University, 197101 Saint Petersburg, Russia}

\author{Nikita A. Olekhno}
\affiliation{School of Physics and Engineering, ITMO University, 197101 Saint Petersburg, Russia}

\date{\today}

\begin{abstract} 
Photonic higher-order topological insulators (HOTI) are characterized by a hierarchy of topologically-protected states with different dimensionalities, making them especially interesting for potential applications that combine strong localization of electromagnetic fields and their robust waveguiding. However, their practical implementation often requires expensive processing techniques and is limited by accessible material parameters. In this paper, we demonstrate the implementation of edge and corner states at radio frequencies in a metasurface composed of split-ring resonators with couplings between dipole and quadrupole modes. We verify, by numerical simulations and experimentally at frequencies of $1.5-1.7$~GHz, that a proposed metasurface supports corner- and edge-localized states in the bulk bandgap. Our results reveal a scalable and easily reconfigurable GHz-range platform that employs printed circuit board technology, thus making crucial steps required for further experimental studies of photonic HOTI and the development of their microwave applications.
\end{abstract}

\maketitle

\section{Introduction}
\label{sec:Introduction}

Photonic topological insulators (PTIs)~\cite{2008_Haldane} are photonic crystals with a band gap that hosts surface-, edge- or corner-localized excitations governed by bulk-boundary correspondence and protected by certain symmetries~\cite{2019_Ozawa}. Such topological states demonstrate an increased resilience to certain types~\cite{2026_Leykam} of structural defects in the PTI structure, e.g., a suppressed back reflection~\cite{2009_Wang} for chiral edge states allowing them to propagate along sharp bends~\cite{2018_Gao} or localization at edges of arbitrary shape for time-reversal invariant phases~\cite{2013_Khanikaev}. These properties of PTIs have drawn constant attempts for the development of their practical applications such as optical fibers~\cite{2025_Zhu} and photonic chips~\cite{2022_Wang, 2024_On, 2024_Tang} at optical frequencies, as well as beamforming chips~\cite{2024_Wang} and integrated antennas~\cite{2023_Jia} for 6G in the THz range. However, the implementation of optical and THz structures requires expensive techniques such as lithography and laser writing, whereas radio-frequency (RF) implementations of PTIs are usually more accessible for experimental studies and production. Examples of RF applications include wireless power transfer~\cite{2021_Song}, signal transmission~\cite{2023_Feis}, and magnetic resonance imaging~\cite{2023_Puchnin} for one-dimensional (1D) arrays~\cite{2024_Guo}, as well as antennas~\cite{2020_Lumer, 2024_Abtahi}, 5G beamformers~\cite{2022_Nagulu}, and even smart clothing with integrated sensor networks~\cite{2026_Li} for two-dimensional (2D) topological metasurfaces.

Recently introduced multipole higher-order topological insulators (HOTIs)~\cite{2017_Benalcazar_Science} demonstrate a rich hierarchy of surface, hinge, and corner states, potentially allowing to combine antennas, waveguides, and resonators in a single structure. However, their experimental realization is complex due to the need to implement couplings between lattice sites with phases shifted by $\pi$. Photonic quadrupole HOTI were implemented at infrared frequencies as 2D microring resonator arrays~\cite{2019_Mittal} and in the RF range with 2D microstrip circuits~\cite{2018_Peterson}. There are also examples of 2D photonic HOTIs governed by rotational symmetries~\cite{2019_Xie, 2019_Yang, 2023_Bobylev} and recent implementations of three-dimensional (3D) photonic HOTIs~\cite{2025_Wang}.

\begin{figure}[b]
  \centering
  \includegraphics[width=8.5cm]{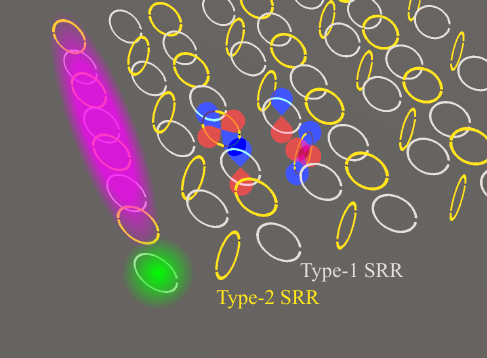}
  \caption{The proposed scheme for implementing a radio-frequency metasurface with $p$-$d$ orbital coupling composed of split-ring resonators (SRRs) of two types: Type-1 (light-gray) featuring a dipole response and Type-2 (yellow) characterized by a quadrupole field pattern, as shown for the unit cell in the center. The regions colored in purple and green highlight edge and corner states, respectively.}
  \label{fig:System}
\end{figure}

\begin{figure*}[tbp]
  \centering
  \includegraphics[width=17cm]{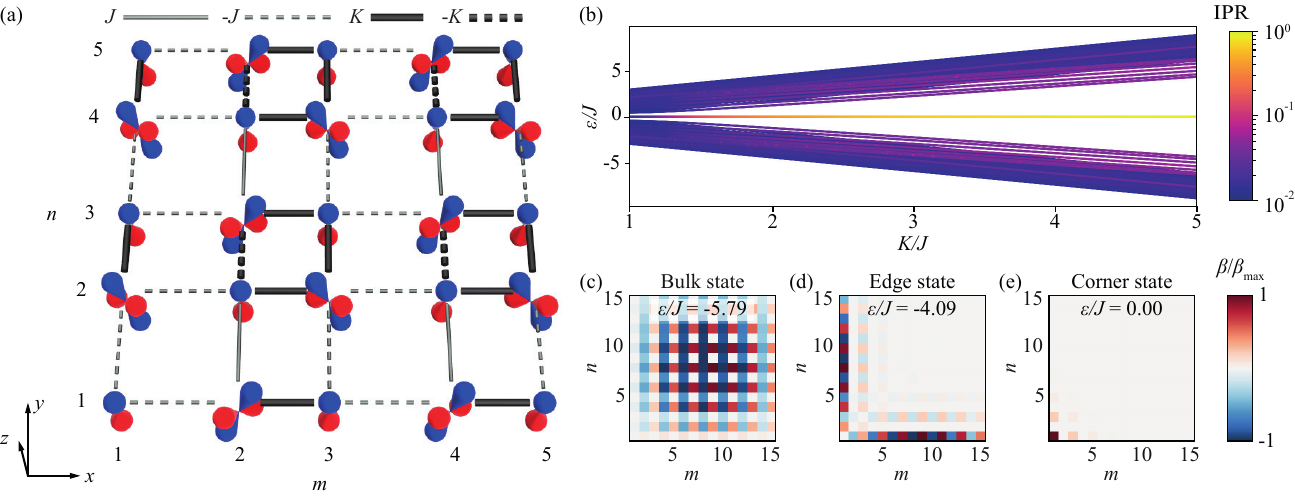}
  \caption{Eigenstates of the tight-binding model. (a) Theoretical model showing the couplings $\pm J$ and $\pm K$ between the nearest resonators in a square lattice of $p$ and $d$ orbitals in the $(xy)$ plane. (b) Spectra of normalized eigenenergies $\varepsilon/J$ as a function of couplings ratio $K/J$ for the tight-binding model with $15 \times 15$ sites. The color shows the values of IPR Eq.~\eqref{eq:IPR} for the corresponding eigenmodes. (c)-(e) Normalized eigenmode profiles $\beta_{n,m}$ corresponding to the (c) bulk, (d) edge, and (e) corner states for the system with $15 \times 15$ sites and couplings $K/J = 5$.}
  \label{fig:Theory}
\end{figure*}

A strategy to obtain quadrupole HOTIs with a synthetic $\pi$-flux by coupling monopole $s$ and dipole $p$ on-site orbitals has recently been introduced~\cite{2022_Schulz} and experimentally realized in a 3D structure with an array of optical waveguides forming a 2D lattice~\cite{2022_Schulz} and hybrid metal-dielectric photonic crystals supporting confined Mie resonances in the GHz range~\cite{2025_Huang}. However, despite clearly demonstrating photonic quadrupole HOTIs in experiment, these structures rely on complex fabrication techniques.

In this paper, we consider a metasurface composed of split-ring resonators (SRRs) with the couplings between dipole and quadrupole modes that hosts edge and corner states, Fig.~\ref{fig:System}. The proposed approach is inexpensive and feasible for the production of large-scale, three-dimensional, and reconfigurable structures, as well as the introduction of nonlinearities, rendering it prospective for future experimental studies of complex PTIs and potential radio-frequency applications. We experimentally realize such a metasurface at GHz frequencies of the L-band mobile-satellite service.

The paper is organized as follows. In Section~\ref{sec:Theory}, we consider properties of the tight-binding model corresponding to the coupling of dipole and quadrupole on-site modes in the nearest-neighbor approximation. Then, in Section~\ref{sec:Simulations}, we numerically simulate the modes of individual split-ring resonators with dipolar and quadrupolar response, the band structure of an infinite SRR lattice, and the eigenmodes of finite systems with different boundaries, confirming the existence of edge- and corner-localized states. The experimental implementation of the proposed structure is discussed in Section~\ref{sec:Experiments}. Section~\ref{sec:Discussion} includes a discussion and an outlook.

\begin{figure*}[tbp]
  \centering
  \includegraphics[width=17cm]{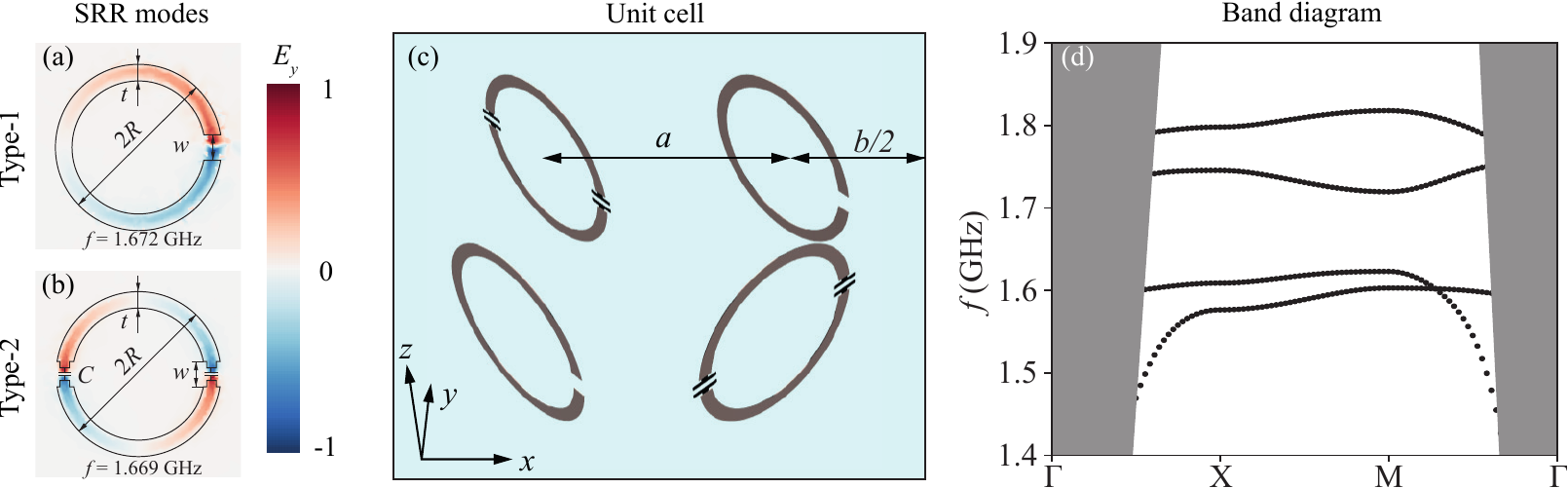}
  \caption{Numerical simulations of the unit cell. (a) Schematics of a Type-1 SRR characterized by the outer radius $R$, ring width $t$, and a gap $w$ superimposed with a numerically simulated distribution of out-of-plane electric field component $E_y$ at the frequency {$f = 1.67$~GHz} at height $0.5$~mm for $R=13.5$~mm, $t=1$~mm, and $w=1.5$~mm. (b) The same for a Type-2 SRR with two gaps loaded by capacitors $C$ that are attached to the pads with sizes $0.6 \times 0.4$~mm$^{2}$ for the parameters $R=13.5$~mm, $t=1$~mm, $w=1.3$~mm, and $C=0.2$~pF. (c) The unit cell used for band structure simulations with the SRRs from (a)-(b) arranged in a dimerized square lattice with periods $a=28$~mm and $b=23$~mm. (d) Band structure obtained for the unit cell model (c) with Floquet boundary conditions. Gray regions correspond to the light cone.}
  \label{fig:SRR_eigenmodes}
\end{figure*}

\section{Properties of the tight-binding model in the nearest-neighbor approximation}
\label{sec:Theory}

We start with considering a dimerized square lattice of $N \times N$ sites in the $(xy)$ plane that hosts dipolar $p$ and quadrupolar $d$ atomic orbitals that are centered at height $z = 0$, Fig.~\ref{fig:Theory}(a). The considered orbital orientations result in phase-alternating effective tunneling couplings $\pm J$ and $\pm K$ between the nearest sites, thus implementing the model characterized by a synthetic $\pi$-flux in each of the four-site plaquettes, analogously to recent models with $s$ and $p$ orbitals implemented in 3D optical waveguide arrays~\cite{2022_Schulz} and $s$, $p$, and $d$ orbitals realized in a 2D RF photonic crystal~\cite{2025_Huang}. In particular, positive couplings $J$ and $K$ are observed between in-phase orbitals, while negative couplings $-J$ and $-K$ emerge between counter-phase orbitals. The considered tight-binding model describes a quadrupole topological insulator~\cite{2022_Schulz} similarly to the Benalcazar-Bernevig-Hughes (BBH) model~\cite{2017_Benalcazar_Science, 2017_Benalcazar_PRB}. The equations describing the tight-binding model are discussed in the Supplementary Material.

To quantify the localization of the eigenstates, we evaluate their inverse participation ratios ($\rm IPRs$)~\cite{1974_Thouless}
\begin{equation}
    {\rm IPR} = \sum_{n,m=1}^{N}|\beta_{n,m}|^4,
    \label{eq:IPR}
\end{equation}
where the summation is performed over all sites of the lattice and the eigenfunctions $\beta$ are normalized to $\sum_{n,m}|\beta_{n,m}|^{2}=1$. For strongly localized states with a non-zero amplitude mostly at a single site, $\rm IPR \to 1$, while for delocalized states $\rm IPR$ tends to $1/N^2$. The eigenenergy spectrum of the model with $15 \times 15$ sites as a function of the coupling ratio $K/J$ is shown in Fig.~\ref{fig:Theory}(b). It is seen that a band gap emerges in the presence of coupling dimerization, and the width of the band gap is proportional to the ratio $K/J$. The eigenmode profiles for the model with $K/J = 5$ are shown in Fig.~\ref{fig:Theory}(c)-(e). Three types of eigenmodes are observed that are characterized by significantly different IPR values: 2D delocalized bulk modes with ${\rm IPR} \propto 1/N^2 \sim 10^{-2}$, Fig.~\ref{fig:Theory}(c), 1D edge states with ${\rm IPR} \propto 1/N \sim 10^{-1}$, Fig.~\ref{fig:Theory}(d), and 0D corner states with ${\rm IPR} \sim 10^{0}$, shown in Fig.~\ref{fig:Theory}(e). The observed edge and corner modes in the considered model~\cite{2022_Schulz} localize at the boundaries with weak couplings $J$ and have staggered amplitude patterns decaying in the bulk. The corner state is pinned to zero energy due to chiral symmetry. The localization of the corner states, as well as the gapped edge states with energies closer to the bulk bands, increases with band gap width.

\section{Numerical simulations}
\label{sec:Simulations}

We proceed with the core idea of realizing dipolar and quadrupolar on-site responses with two types of SRRs having the same diameter $2R$ and ring width $t$. In particular, we consider an SRR with a single gap of width $w$ (Type-1), Fig.~\ref{fig:SRR_eigenmodes}(a), and a two-gap SRR with gaps of width $w$ shunted by the capacitors $C$ (Type-2), Fig.~\ref{fig:SRR_eigenmodes}(b). The gaps in a Type-2 SRR incorporate additional pads for the placement of capacitors. First, we perform numerical simulations of eigenmodes for individual SRRs made of a perfect electric conductor of zero thickness using Eigenmode solver in CST Microwave Studio. The capacitors $C$ in a Type-2 SRR allow to adjust the frequency of its quadrupolar mode to the same frequency range as a Type-1 SRR displays dipolar response, as illustrated in Fig.~\ref{fig:SRR_eigenmodes}(a),(b) for $f=1.67$~GHz. Such a quadrupolar mode of a Type-2 SRR results from the anti-symmetric combination of dipolar modes of two half-rings.

Next, we consider numerically the bulk band structure for an infinite lattice with a unit cell shown in Fig.~\ref{fig:SRR_eigenmodes}(c). The unit cell consists of two parallel Type-1 SRRs and two orthogonally-oriented Type-2 SRRs placed in the corners of a square with the intra-cell distance $a=28$~mm and the inter-cell distance $b=23$~mm. The gaps of all SRRs are located in the same $(xy)$ plane. We perform simulations with Floquet boundary conditions for the electric and magnetic fields. As shown in Fig.~\ref{fig:SRR_eigenmodes}(d) for the high-symmetry point trajectory $\Gamma(0,0)-X(\pi,0)-M(\pi,\pi)-\Gamma(0,0)$ in the Brillouin zone $(k_x,k_y)$, where $k_{x}$ and $k_{y}$ are the wave numbers in the directions $x$ and $y$, respectively, such a metasurface possesses a bulk band gap at frequencies $1.62-1.72$~GHz.

\begin{figure*}[tbp]
  \centering
  \includegraphics[width=17cm]{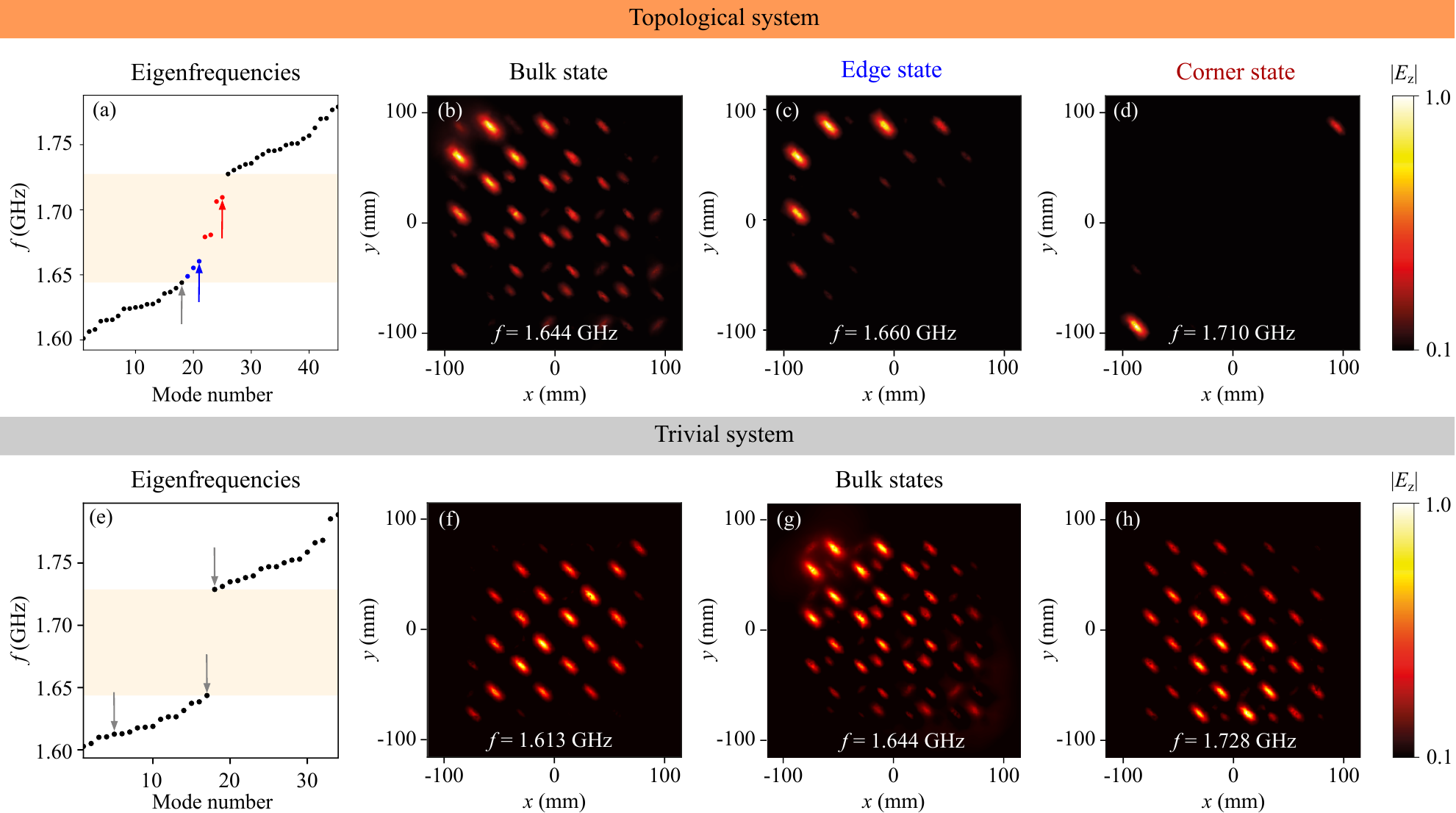}
  \caption{Numerical simulations of finite structures with different boundaries. (a) Eigenfrequency spectrum for the finite lattice of $8 \times 8$ SRRs with periods $a = 28$~mm and $b = 23$~mm with weak couplings at the boundaries. Orange shading corresponds to the bulk band gap. (b)-(d) Numerically calculated distributions of the out-of-plane electric field component $E_z$ for the eigenmodes corresponding to (b) bulk ($f = 1.644$~GHz, ${\rm IPR}=2.0 \cdot 10^{-4}$), (c) edge ($f = 1.660$~GHz, ${\rm IPR}=8.4 \cdot 10^{-4}$), and (d) corner ($f = 1.710$~GHz, ${\rm IPR}=30.7 \cdot 10^{-4}$) states, respectively. (e) Eigenfrequency spectrum for the finite lattice of $8 \times 8$ SRRs with periods $a = 28$~mm and $b = 23$~mm with strong couplings at the boundaries. (f)-(h) Numerically calculated distributions of $E_z$ for the bulk modes corresponding to eigenfrequencies (f) $f = 1.613$~GHz, (g) $f = 1.644$~GHz, and (h) $f = 1.728$~GHz. Arrows in (a) and (e) denote the frequencies of the plotted eigenmodes.}
  \label{fig:Simulations}
\end{figure*}

To study whether this band gap hosts edge and corner states, we proceed with numerical simulations of the eigenmodes of a finite system of size $8 \times 8$ resonators with weak links $a=28$~mm at the boundaries. The eigenmode spectrum is shown in Fig.~\ref{fig:Simulations}(a) and demonstrates two bulk bands and a band gap at frequencies $1.644-1.728$~GHz. Examples of eigenmodes are shown in Fig.~\ref{fig:Simulations}(b)-(d) considering the out-of-plane component of the electric field $E_{z}$ that is expressed primarily in dipolar Type-1 SRRs. It is seen that the first eigenmode below the band gap indeed represents a bulk state, Fig.~\ref{fig:Simulations}(b). In a band gap, it is followed by gapped edge states localized at the weak-link edges with the example shown in Fig.~\ref{fig:Simulations}(c). Finally, there are four corner states with energies in the band gap that demonstrate strong localization at the corners with weak links, Fig.~\ref{fig:Simulations}(d).

Next, we numerically simulate the system consisting of $8 \times 8$ SRRs with the same periods $a = 28$~mm and $b = 23$~mm, but with links $b$ at the boundaries, corresponding to strong links $K$ in the model of Fig.~\ref{fig:Theory}(a). As theory predicts~\cite{2017_Benalcazar_Science,2017_Benalcazar_PRB}, such a system is topologically trivial and should not host any edge or corner states in the band gap. The numerical results in Fig.~\ref{fig:Simulations} fully confirm this intuition, demonstrating the presence of a band gap in Fig.~\ref{fig:Simulations}(e) without any in-gap states. As can be directly verified, all eigenmodes of such a system represent delocalized bulk modes, with examples of the eigenmodes at the edges of the band gap and the middle of a bulk band shown in Fig.~\ref{fig:Simulations}(f)-(h).

The obtained numerical results qualitatively agree with the tight-binding model in Fig.~\ref{fig:Theory}(a) despite the apparent presence of long-range couplings between the resonators. Breaking of exact chiral symmetry is observed from the spectra in Fig.~\ref{fig:Simulations}(a),(e), which is typical for photonic structures~\cite{2025_Wang, 2025_Huang} and is related to a linear dispersion of the lower bands at the $\Gamma$-point, radiative and Ohmic losses, and the presence of couplings between the next-nearest resonators~\cite{2022_D4}. In particular, the upper branch of edge states mixes with bulk continuum, and the spectral positions of corner states are shifted from the middle of the band gap.

Similarly to the optical system of Schulz and coauthors~\cite{2022_Schulz}, the unit cell of the considered model shown in Fig.~\ref{fig:SRR_eigenmodes}(c) lacks rotational symmetries. As a result, the considered system cannot represent a $C_4$-symmetric topological crystalline insulator~\cite{2019_Benalcazar}, in contrast to several demonstrations of photonic HOTI in 2D dielectric structures~\cite{2019_Xie, 2019_Chen} and SRR arrays~\cite{2023_Bobylev}. The comparison of unit cells with even and odd numbers of effective negative couplings qualitatively matches the behavior of the BBH model~\cite{2017_Benalcazar_PRB, 2022_Schulz} (Fig.~S4 in the Supplementary Material), pointing to the presence of an effective $\pi$-flux. However, the considered model features only a single reflection symmetry with respect to one diagonal of the structure, while quadrupole moment quantization in BBH-type models in the absence of chiral symmetry requires two non-commuting reflection symmetries~\cite{2017_Benalcazar_Science, 2017_Benalcazar_PRB}. Hence, the constructed system cannot represent a quadrupole insulator as well.

To demonstrate topological origin~\cite{2021_Khalaf} of the corner states in the considered SRR array, we study a set of extended tight-binding models as discussed in Section~S4 in the Supplementary Material. In particular, we consequently add diagonal couplings between various Type-1 resonators within the unit cell and introduce an additional dimerization of inter-cell and intra-cell couplings associated with finite size of the resonators, thus breaking chiral symmetry, inversion, and various reflection symmetries. As shown in Section~S5 of the Supplementary Material, bulk and edge band gaps remain open and in-gap corner states are observed upon the adiabatic introduction of such modifications in the initial BBH-type tight-binding model, see Fig.~S7. Moreover, a direct calculation~\cite{2017_Benalcazar_PRB} reveals fractional corner charges~\cite{2020_He, 2020_Peterson, 2023_Liang} in all of the considered extended models, with charges at the two symmetric corners with coordinates $(1,1)$ and $(N,N)$ remaining close to the value $Q=1/2$ for the BBH model.

\section{Experimental demonstration}
\label{sec:Experiments}

\begin{figure}[tbp]
  \centering
  \includegraphics[width=8cm]{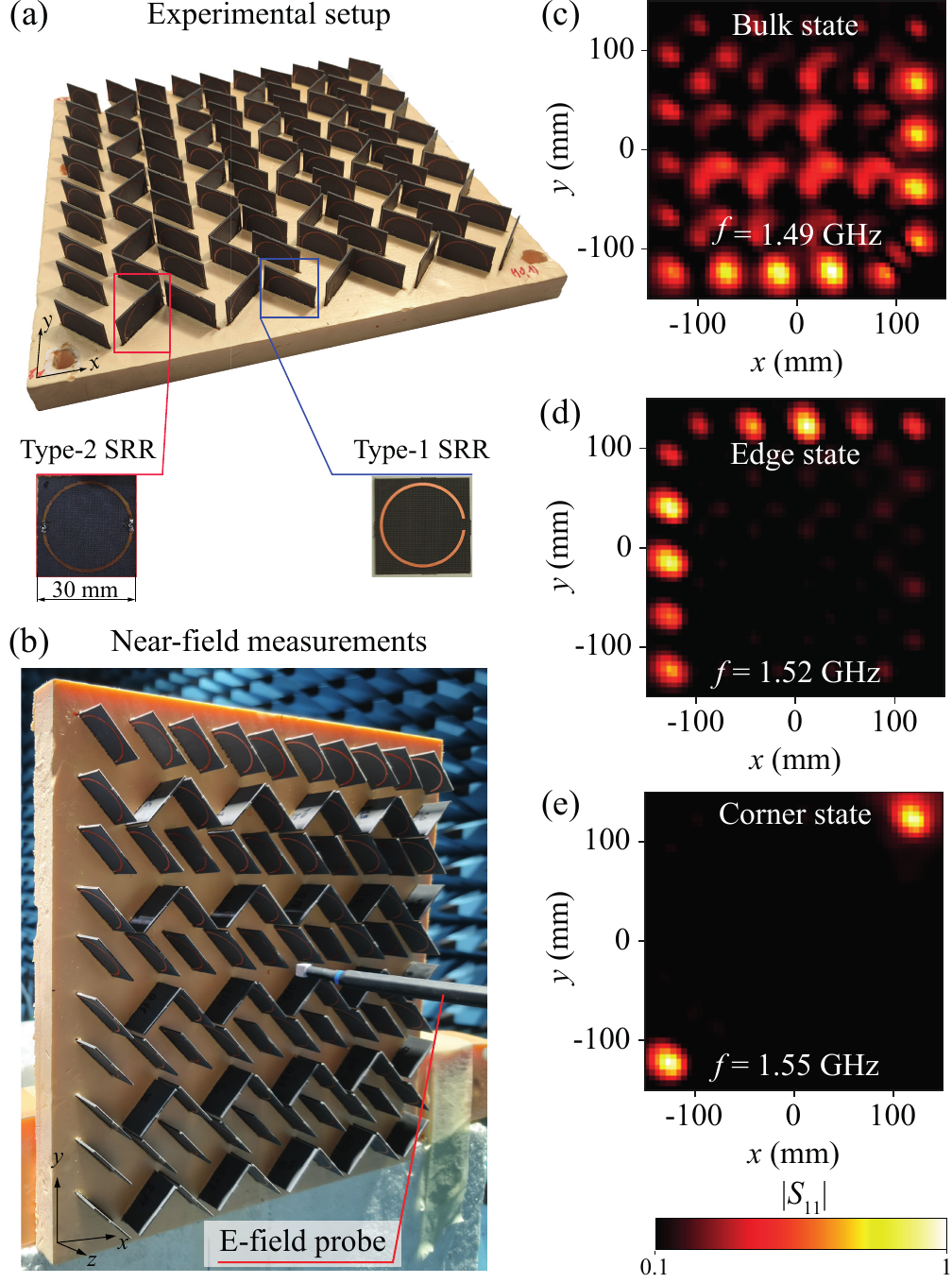}
  \caption{Experimental studies. (a) Experimentally realized metasurface with an array of $10 \times 10$ printed circuit board SRRs arranged in a dimerized square lattice with periods $a = 28$~mm and $b = 23$~mm. The insets demonstrate Type-1 and Type-2 SRRs. (b) The setup including E-field probe that is moved by a spatial scanner. (c)-(e) Experimentally measured distributions of the reflection coefficient $|S_{11}|$, which characterizes the near-field distribution of the electric field component $E_z$, corresponding to (c) bulk, (d) edge, and (e) corner states. Each distribution is normalized to its maximum.}
  \label{fig:Experiments}
\end{figure}

To verify the results experimentally, we constructed an array of $10 \times 10$ SRRs implemented as printed circuit boards (PCBs) with a copper layer on a $1.016$~mm thick F4BM255 dielectric substrate with in-plane dimensions of $30 \times 30$~mm, Fig.~\ref{fig:Experiments}(a). The SRRs have parameters $R=13.5$~mm, $t=1$~mm, $w=3.5$~mm (Type-1) or $w=1.3$~mm (Type-2), providing a close matching of the dipole and quadrupole eigenmodes of Type-1 and Type-2 SRRs at frequencies $f = 1.559$~GHz and $f=1.565$~GHz, respectively. Such parameters are obtained by numerical simulations of SRRs taking into account a dielectric substrate, as discussed in the Supplementary Material. Type-2 SRRs are loaded with a series connection of two capacitors with values of $0.3$~pF and $0.6$~pF, which results in a total capacitance $C=0.2$~pF. The capacitors are soldered to $0.6 \times 0.4$~mm$^2$ pads. Resonators are arranged in a dimerized square lattice with periods $a = 28$~mm and $b = 23$~mm using an extruded polystyrene foam (XPS) substrate, as shown in Fig.~\ref{fig:Experiments}(a). To study experimentally the electromagnetic response of the considered system to a near-field excitation, we measure the reflection coefficient $S_{11}$ using E-field probe XF-E~04s, that detects the vertical component of the electric field $E_z$. Measurements were performed with the help of the PLANAR C4490 vector network analyzer. During the measurements, the probe was automatically moved in the plane of the lattice at a distance of $5$~mm from the edge of the split-ring PCBs using the Trim TMC~3113 spatial scanner, Fig.~\ref{fig:Experiments}(b). The scanning step along the $x$- and $y$-axes is $5$~mm.

Examples of measured S-parameter distributions are shown in Fig.~\ref{fig:Experiments}(c)-(e). In particular, Fig.~\ref{fig:Experiments}(c) represents a bulk state for frequency $f = 1.49$~GHz, featuring a delocalization of the electric field (${\rm IPR} = 1.2 \cdot 10^{-3}$). As the frequency increases to the bulk band gap, such a delocalized pattern is replaced by an edge state with a pronounced localization at the weak-link boundaries (left and top), as shown in Fig.~\ref{fig:Experiments}(d) for $f=1.52$~GHz (${\rm IPR} = 3.1 \cdot 10^{-3}$). Finally, the dimensionality of this edge state reduces further upon increasing the frequency, and a corner localization emerges, as displayed by the profile corresponding to $f=1.55$~GHz (${\rm IPR} = 7.0\cdot 10^{-3}$) in Fig.~\ref{fig:Experiments}(e). The degrees of localization for the experimentally measured distributions agree well with those obtained in numerical simulations.

Notably, the experimentally obtained S-parameter maps in Fig.~\ref{fig:Experiments}(c)-(e) demonstrate an excellent agreement with numerical simulations of the eigenmodes in Fig.~\ref{fig:Simulations}(b)-(d) even in the presence of a considerable off-diagonal (in couplings between the sites) and diagonal (in on-site energies) disorder related to fluctuations in the resonant frequencies of Type-2 SRRs that reach $18$~MHz in the experimental setup (see Fig.~S9 in the Supplementary Material). In particular, quadrupole mode frequencies of individual Type-2 SRRs vary from $1.553$~GHz to $1.571$~GHz with a mean frequency of $1.563$~GHz, while a mean frequency of the dipole mode of Type-1 SRRs is $1.564$~GHz and individual frequencies are distributed from $1.563$~GHz to $1.566$~GHz. The reason for such high fluctuations is the $0.05$~pF tolerance of the capacitors, which were intentionally not sorted to assess the disorder robustness of the setup. The measurements of absorption spectrum for all sites of the resonator array are discussed in the Supplementary Material.

\section{Conclusion and Outlook}
\label{sec:Discussion}

The obtained numerical and experimental results can be further extended by considering long-range couplings between the SRRs in the theoretical model. Such couplings are shown to be important for the description of topological states in bianisotropic systems~\cite{2025_Rozenblit, 2026_Rozenblit}, can induce HOTI phases in ordinary topological insulators~\cite{2022_D4,2026_Gutierrez_Llorente} and result in additional types of corner states~\cite{2020_Li}. For example, as discussed in the Supplementary Material, taking into account the couplings between diagonally-opposite Type-1 resonators in the unit cell results in the asymmetry of the spectrum and the splitting of corner state frequencies that are also seen in numerical simulations and experiments. Several long-range generalizations of the BBH model have recently been considered~\cite{2020_Li_Generalized, 2022_Benalcazar, 2023_Yang}. Systems such as fully coupled lattices of SRRs with dipolar interactions~\cite{2023_Luo} and dielectric resonators (see the Supplementary Material for Ref.~\cite{2020_Li}) have also been considered theoretically.

From the applied perspective, the demonstrated approach of combining dipolar and quadrupolar excitations of SRRs to implement PTIs at radio-frequencies resolves many limitations characteristic for dielectric resonators, e.g., bonds for achievable permittivity and geometric restrictions of fabrication techniques. Indeed, one can easily adjust the geometry of PCB strip lines that form SRRs and load such SRRs with different lumped elements. For example, along with capacitors and inductors, one may include varactors, light-emitting diodes, and photodiodes to implement nonlinearities~\cite{2012_Slobozhanyuk, 2018_Dobrykh, 2018_Hadad}, which we envision as a prospective direction for further studies. Moreover, the proposed scheme can be applied to implement 3D higher-order PTIs~\cite{2025_Wang, 2025_Liu, 2025_Lai} or to construct metasurfaces consisting of several domains. Finally, the proposed design with a set of individual PCBs can be applied to implement mechanically reconfigurable structures~\cite{2019_Wang, 2022_Xu, 2025_Qin}.

\textit{Acknowledgements.} We are grateful to Anton Nalitov for stimulating discussions and thank Alexey Dmitriev, Mikhail Buzakov, and Alexander Berdnikov for assistance in preparation of the experimental setup and measurements. The work was supported by the Russian Science Foundation (project~25-22-00768).


%

\end{document}


\title{Supplementary Material\\~\\Topological corner states in metasurfaces based on split-ring resonators with dipole-quadrupole couplings}

\author{Alina~D.~Rozenblit}
\email{alina.rozenblit@metalab.ifmo.ru}
\affiliation{School of Physics and Engineering, ITMO University, 49 Kronverksky pr., bldg. A, 197101 Saint Petersburg, Russia}

\author{Nikita~A.~Olekhno}
\affiliation{School of Physics and Engineering, ITMO University, 49 Kronverksky pr., bldg. A, 197101 Saint Petersburg, Russia}

\date{\today}

\maketitle

\tableofcontents

\newpage

\onehalfspacing

\section{Tight-binding model}
\label{sec:TB}

\begin{figure}[b]
  \includegraphics[width=8.5cm]{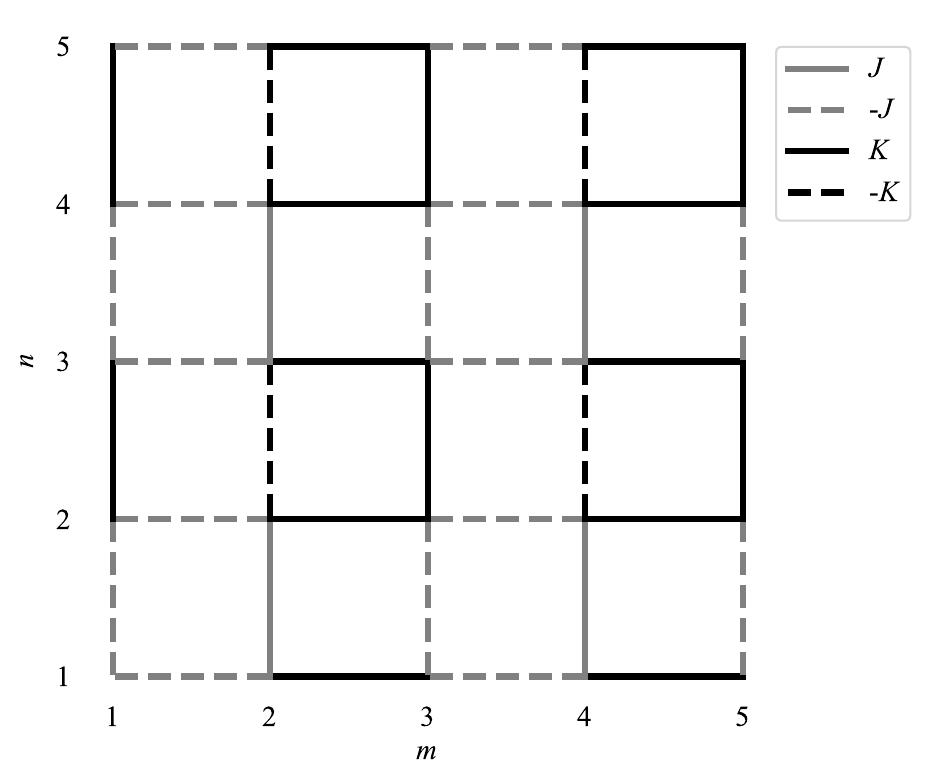}
  \caption{The example of the considered model with $5 \times 5$ sites coupled through tunneling couplings $\pm J$ and $\pm K$.}
  \label{fig:Tight_binding}
\end{figure}

The tight-binding model having the form of a lattice with $N \times N$ sites that we discuss in the main text is depicted in Fig.~\ref{fig:Tight_binding}. For odd $N$, its eigenstates are described by the following tight-binding equations for the complete set of non-equivalent sites:
\begin{equation}
    \begin{cases}
        \varepsilon \beta_{2n,2m} = - K(\beta_{2n, 2m+1} - \beta_{2n+1, 2m}) - J(\beta_{2n-1, 2m} - \beta_{2n, 2m-1}),
        \\
        \varepsilon \beta_{2n,2m+1} = - K(\beta_{2n+1, 2m+1} + \beta_{2n, 2m}) + J(\beta_{2n, 2m+2} + \beta_{2n-1, 2m+1}),
        \\
        \varepsilon \beta_{2n+1,2m} = - K(\beta_{2n+1, 2m+1} - \beta_{2n, 2m}) - J(\beta_{2n+2, 2m} - \beta_{2n+1, 2m-1}),
        \\
        \varepsilon \beta_{2n+1,2m+1} = - K(\beta_{2n+1, 2m} + \beta_{2n, 2m+1}) + J(\beta_{2n+1, 2m+2} + \beta_{2n+2, 2m+1}),
        \\
        \varepsilon \beta_{1,2m+1} = - K\beta_{1, 2m} + J(\beta_{1, 2m+2} + \beta_{2, 2m+1}),
        \\
        \varepsilon \beta_{1,2m} = - K\beta_{1, 2m+1} - J(\beta_{2, 2m} - \beta_{1, 2m-1}),
        \\
        \varepsilon \beta_{2n,1} = - K\beta_{2n+1, 1} + J(\beta_{2n, 2} + \beta_{2n-1, 1}),
        \\
        \varepsilon \beta_{2n+1,1} = - K\beta_{2n, 1} +J(\beta_{2n+1, 2} + \beta_{2n+2, 1}),
        \\
        \varepsilon \beta_{N,2m} = - K(\beta_{N, 2m+1} - \beta_{N-1, 2m}) - J\beta_{N, 2m-1},
        \\
        \varepsilon \beta_{N,2m+1} = - K(\beta_{N, 2m} + \beta_{N-1, 2m+1}) - J\beta_{N, 2m+2},
        \\
        \varepsilon \beta_{2n,N} = - K(\beta_{2n, N-1} + \beta_{2n+1, N}) + J\beta_{2n-1, N},
        \\
        \varepsilon \beta_{2n+1,N} = - K(\beta_{2n+1, N-1} + \beta_{2n, N}) + J\beta_{2n+2, N},
        \\
        \varepsilon \beta_{1,1} = J(\beta_{1, 2} + \beta_{2, 1}),
        \\
        \varepsilon \beta_{1,N} = -K\beta_{1, N-1} + J \beta_{2, N},
        \\
        \varepsilon \beta_{N,1} = -K\beta_{N-1, 1} + J \beta_{N, 2},
        \\
        \varepsilon \beta_{N,N} = -K(\beta_{N-1, N} + \beta_{N, N-1}),
    \end{cases}
    \label{eq:Quadrupole_TB}
\end{equation}
where $\varepsilon$ stands for the eigenenergy, $\beta_{n,m}$ is the amplitude of the eigenfunction in the lattice site with coordinates $(n,m)$, and the indices $n$ and $m$ take values in the range $1 \leq n,m \leq (N-3)/2$. In photonic systems, such eigenstates and eigenvalues describe electromagnetic field distributions and resonance frequencies, respectively. The spectra of normalized eigenenergies $\varepsilon/J$ and eigenmodes $\beta_{n,m}$ are obtained by solving the eigenvalue problem Eq.~\eqref{eq:Quadrupole_TB} numerically in Python.

\section{Design and geometry optimization of resonators}
\label{sec:Resonators}

\begin{figure*}[b]
  \includegraphics[width=7.5cm]{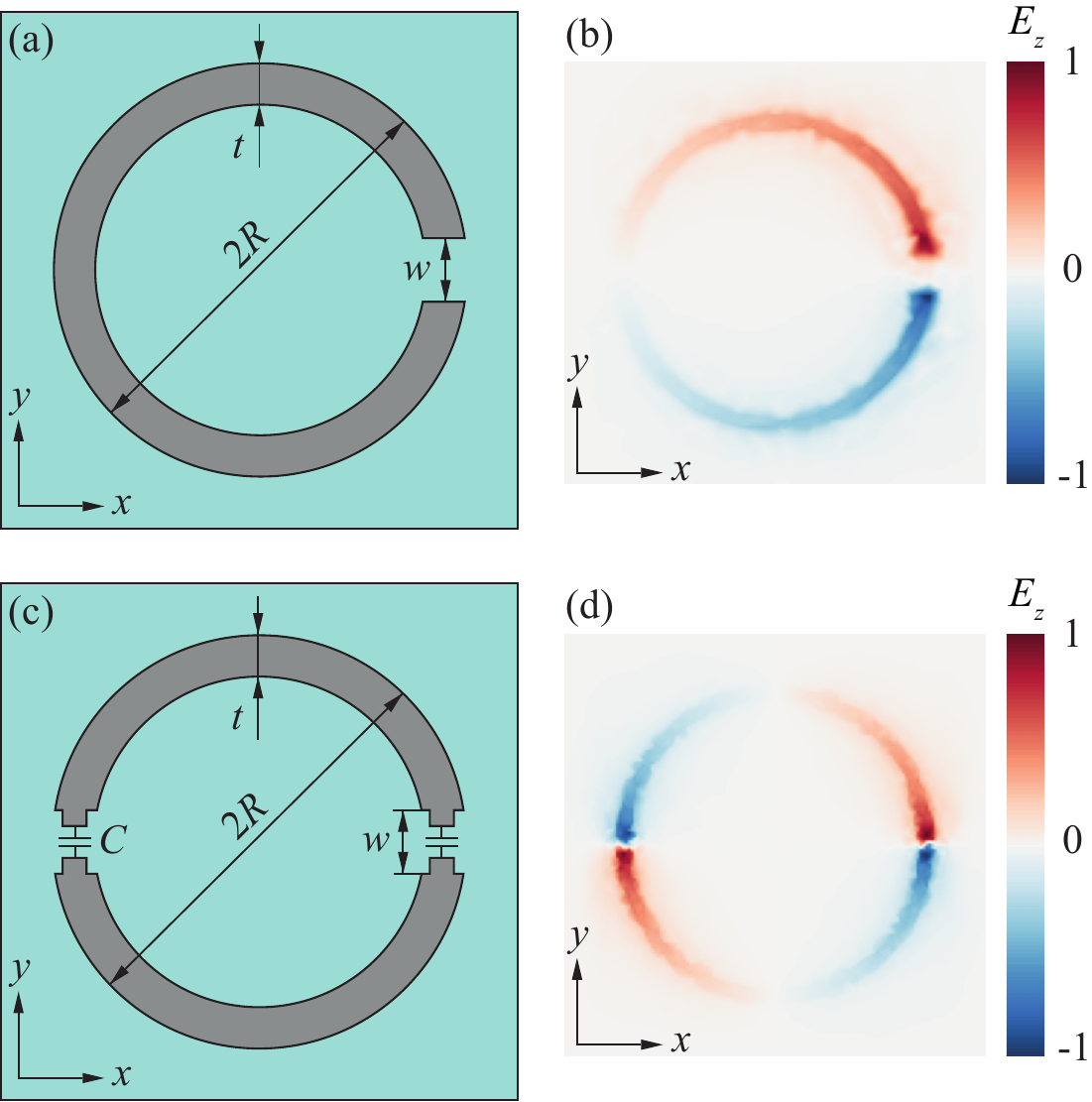}
  \caption{(a) Schematic of a Type-1 SRR (gray) with the outer radius $R$, width $t$, and a gap $w$ placed on the substrate (blue). (b) The distribution of the out-of-plane electric field component $E_z$ at frequency $f = 1.559$~GHz for the parameters $R=13.5$~mm, $t=1$~mm, and $w=3.5$~mm. (c) The same as panel (a), but for Type-2 SRR loaded by capacitors $C$. (d) The distribution of the out-of-plane electric field component $E_z$ at frequency $f = 1.565$~GHz for the parameters $R=13.5$~mm, $t=1$~mm, $w=1.3$~mm, and $C = 0.2$~pF.}
  \label{fig:Realistic_SRR}
\end{figure*}

Prior to experimental realization, we performed geometry optimization for the split-ring resonators (SRRs) placed on a dielectric substrate with the help of CST Microwave Studio. To this end, we introduce a square-shaped dielectric substrate with permittivity $\varepsilon = 2.55$, height $1.016$~mm, and side length $30$~mm in the numerical model, Fig.~\ref{fig:Realistic_SRR}(a),(c). The center of the substrate aligns with the center of the SRR. We also replace the zero-thickness perfect electric conductor by an $18$~$\mu$m copper layer having the conductivity $5.8 \cdot 10^{7}$~S/m in the numerical model for both types of SRRs. In the presence of such a dielectric substrate, the frequency of the fundamental mode supported by Type-2 SRR with the parameters $R=13.5$~mm, $t=1$~mm, $w=1.3$~mm, and $C = 0.2$~pF decreases from $f=1.67$~GHz to $f=1.565$~GHz. The resonance of Type-1 SRR can be adjusted by changing the width of the gap. For example, Type-1 SRR with parameters $R=13.5$~mm, $t=1$~mm, and $w=1.5$~mm placed on the same dielectric substrate has resonance at frequency $f = 1.559$~GHz instead of $f=1.67$~GHz if the width of the gap $w$ is increased from $1.5$ to $3.5$~mm. The introduction of the dielectric substrate preserves the patterns of the dipole and quadrupole eigenmodes of Type-1 and Type-2 SRRs, as shown in Fig.~\ref{fig:Realistic_SRR}(b),(d).

\begin{figure}[tbp]
  \includegraphics[width=12cm]{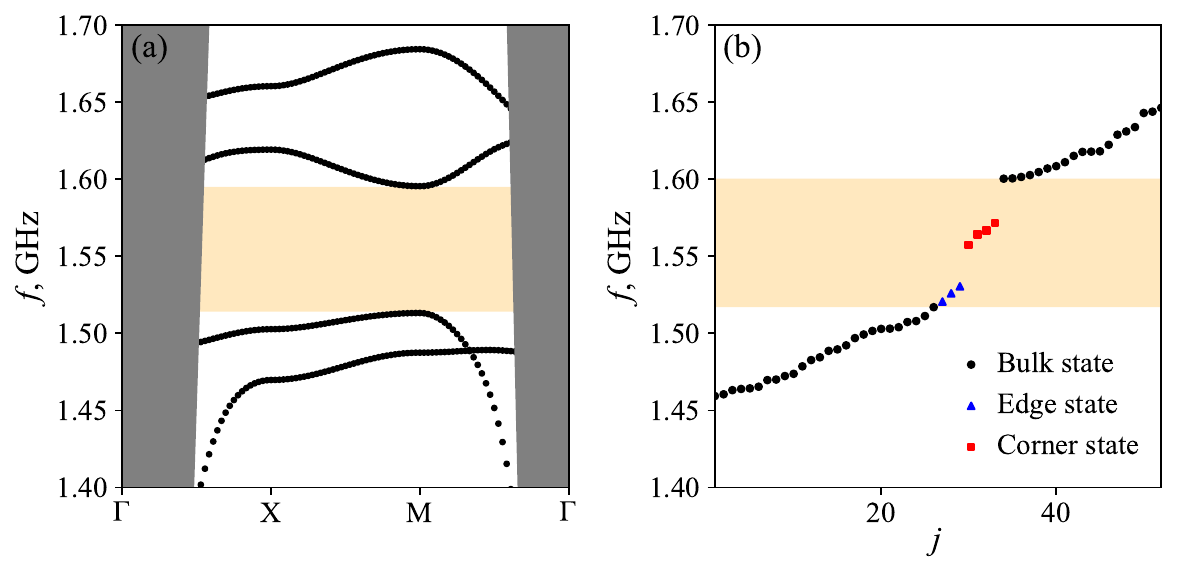}
  \caption{(a) Dispersion diagram for the lattice of SRRs placed at a dielectric substrate with permittivity $\varepsilon=2.55$ and arranged with periods $a = 28$~mm and $b = 23$~mm. Gray regions correspond to the light cone. (b) Eigenfrequency spectra for the array of $8 \times 8$ SRRs at a dielectric substrate with $\varepsilon=2.55$ and periods $a = 28$~mm and $b = 23$~mm. Black, blue, and red markers correspond to bulk, edge, and corner states, respectively. Orange shading shows the bulk band gap.}
  \label{fig:Substrate_dispersion_and_spectra}
\end{figure}

Numerically simulated dispersion diagram for the unit cell with Floquet boundary conditions and the SRRs placed at the dielectric substrate demonstrates a band gap ranging from $1.51$~GHz to $1.60$~GHz, as shown for the lattice with periods $a = 28$~mm and $b = 23$~mm in Fig.~\ref{fig:Substrate_dispersion_and_spectra}(a). Thus, the decrease in the resonance frequencies of Type-1 and Type-2 SRRs on dielectric substrate leads to a downward shift of the band gap frequencies by about $100$~MHz. The same is demonstrated by the eigenfrequencies spectra calculated for an array of $8 \times 8$ SRRs placed at dielectric substrate and arranged in lattice with periods $a = 28$~mm and $b = 23$~mm, Fig.~\ref{fig:Substrate_dispersion_and_spectra}(b). Note, that in the finite array of resonators with the dielectric substrate, the frequencies of the four corner states are closer, compared to those in the system without the substrate shown in Fig.4(a) in the main text.

\section{Modes of the unit cell}
\label{sec:Unit_cell}

\begin{figure*}[b]
  \includegraphics[width=10cm]{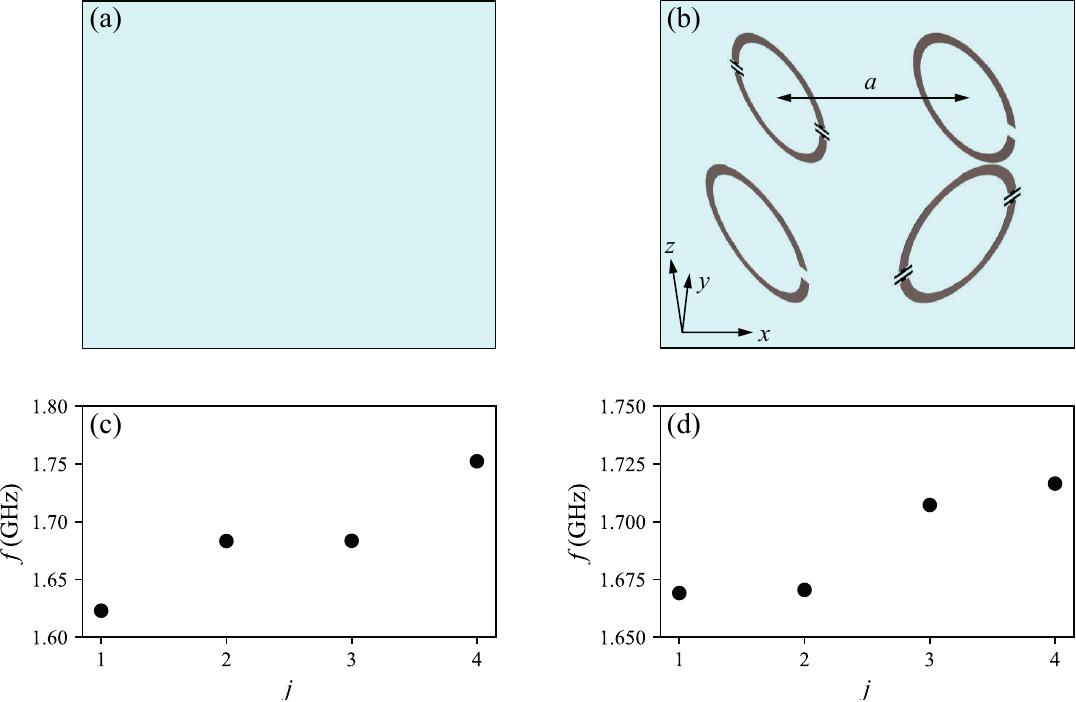}
  \caption{(a) Schematics of the unit cell with $a=28$~mm formed by Type-1 and Type-2 SRRs and featuring two negative couplings. (b) The unit cell with $a=28$~mm and a single negative coupling. (c) The frequencies of eigenmodes of the unit cell (a). (d) The frequencies of eigenmodes of the unit cell (b).}
  \label{fig:Unit_cell}
\end{figure*}

In this Section, we study resonances of individual unit cells. We start with constructing a unit cell with one of Type-2 SRRs being additionally rotated by 90 degrees to obtain two negative $p$-$d$ couplings in the unit cell, as shown in Fig.~\ref{fig:Unit_cell}(a). It is seen that such a unit cell has the same spatial symmetry as the unit cell in Fig.~\ref{fig:Unit_cell}(b) considered in the main text. The distance between the centers of the SRRs for both unit cells is $a=28$~mm, and the parameters of the SRRs are $R=13.5$~mm, $t=1$~mm, and $w=1.5$~mm for Type-1 SRRs and $R=13.5$~mm, $t=1$~mm, $w=1.3$~mm, and $C = 0.2$~pF for Type-2 SRRs.

As seen in Fig.~\ref{fig:Unit_cell}(c), the spectrum for the unit cell with two negative couplings features two approximately degenerate modes at frequencies $f_2=1.6833$~GHz and $f_3=1.6835$~GHz and two non-degenerate modes with frequencies $f_1=1.6229$~GHz and $f_4=1.7523$~GHz, in close analogy to the results for a $2\pi$-flux unit cell in the BBH model~\cite{2017_Benalcazar_PRB, 2022_Schulz} that is gapless at half-filling and features a pair of doubly-degenerate modes with energies $E_{2,3}/K=0$ and two modes with energies $E_{1,4}/K=\pm 2$, where $K$ is the intra-cell coupling. In turn, introducing a single negative coupling results in the formation of two pairs of doubly-degenerate eigenstates shown in Fig.~\ref{fig:Unit_cell}(d). The frequencies of the four eigenmodes are $f_1=1.6691$~GHz, $f_2=1.6705$~GHz, $f_3=1.7072$~GHz, and $f_4=1.7165$~GHz, resembling the formation of two pairs of modes with energies $E_{1,2}/K = -\sqrt{2}$ and $E_{3,4}/K = \sqrt{2}$ in the $\pi$-flux unit cell of the BBH model that is gapped at half-filling and thus can be used in the construction of a quadrupole insulator~\cite{2017_Benalcazar_PRB}. A slight non-degeneracy of $f_{1,2}$ and especially $f_{3,4}$ for the numerically simulated modes is related to additional coupling between diagonally-opposite Type-1 SRRs, the asymmetry in $p$-orbitals of Type-1 SRRs, and a potential difference in coupling magnitudes between Type-1 SRRs and Type-2 SRRs with different orientations.

\section{Extended tight-binding models}
\label{sec:Extended_TB}

\begin{figure}[b]
  \includegraphics[width=15cm]{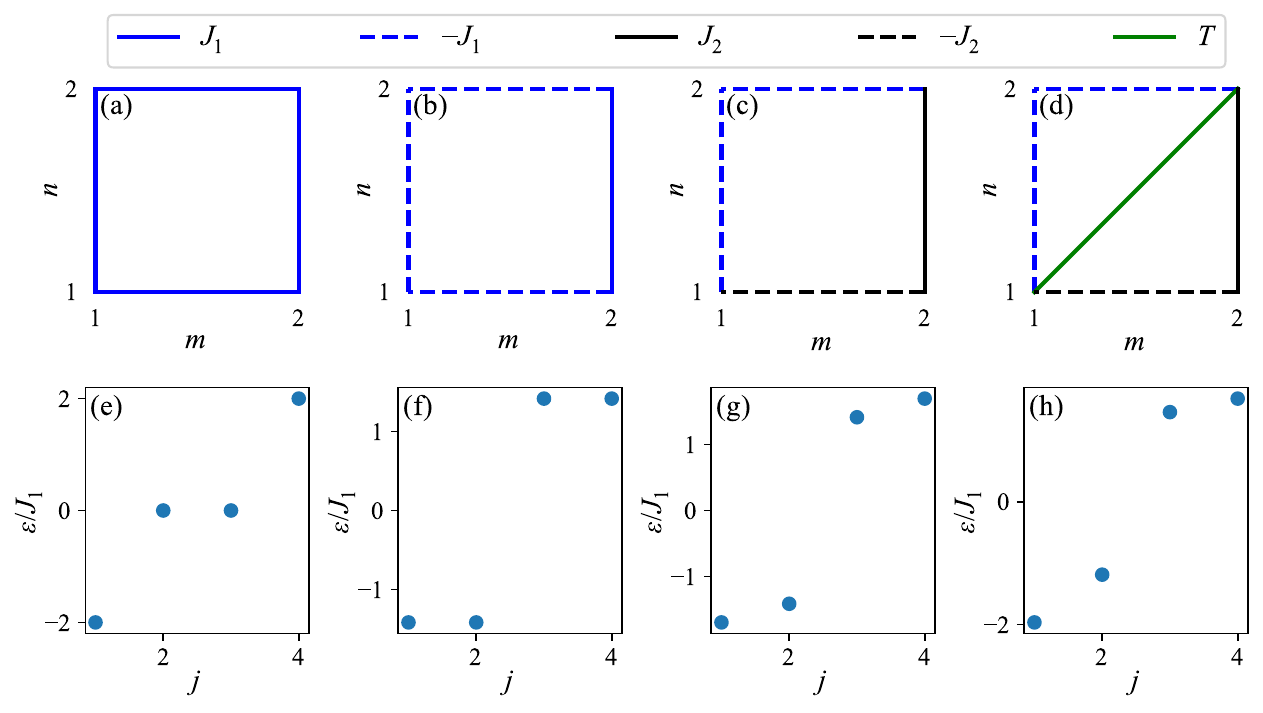}
  \caption{Unit cells of tight-binding models with (a) positive coupling constant $J_1$, (b) coupling constants $\pm J_1$, (c) coupling constants $\pm J_1$ and $\pm J_2$, and (d) coupling constants $\pm J_1$ and $\pm J_2$ and the next-nearest coupling $T$. (e)-(h) Eigenvalues $\varepsilon/J_1$ for the corresponding unit cells in panels (a)-(d) for the following coupling constant values: $J_2=1.2J_{1}$, $T = 0.5J_{1}$.}
  \label{fig:Extended_unit_cell}
\end{figure}

In this Section, we construct tight-binding models with additional modifications relevant to the considered system. Indeed, while the energy levels of the zero-flux unit cell in Fig.~\ref{fig:Extended_unit_cell}(a),(e) demonstrate a good correspondence with numerical simulations in Fig.~\ref{fig:Unit_cell}(a), for the unit cell with a negative coupling in Fig.~\ref{fig:Unit_cell}(b), the splitting in the frequencies of pairs of nearly-degenerate modes is observed which is absent in the idealized model of Fig.~\ref{fig:Extended_unit_cell}(b),(f). Moreover, the spectra in Fig.~4(a),(e) in the main text demonstrate an asymmetry in the edge states, the offset of corner state frequencies from mid gap, and their splitting, that are not observed in the symmetric energy spectrum of the model in Fig.~2. 

First, from the unit cell Fig.~\ref{fig:Unit_cell}(b) it is seen that the couplings between Type-1 SRRs and the top left or the bottom right Type-2 SRRs will be different because of the different orientation of Type-2 SRRs. Then, an obvious modification of the tight-binding model Fig.~\ref{fig:Tight_binding} consists in introducing an additional dimerization in couplings $J$,$K$ and replacing them with couplings $J_{1}$, $J_{2}$ and $K_{1}$, $K_{2}$. As shown in Fig.~\ref{fig:Extended_unit_cell}(c),(e), such a dimerization leads to the symmetric splitting of two degenerate energy levels. In turn, adding a diagonal coupling between two sites corresponding to Type-1 SRRs results in the asymmetry of this splitting observed in Fig.~\ref{fig:Extended_unit_cell}(d),(h) which resembles the one seen in numerical simulations of the unit cell in Fig.~\ref{fig:Unit_cell}(b). This diagonal coupling is the only additional coupling which should be taken into account, since the interaction between two Type-2 SRRs along the orthogonal diagonal of the unit cell zeroes out due to symmetry of the SRR modes.

\begin{figure}[tbp]
  \includegraphics[width=16cm]{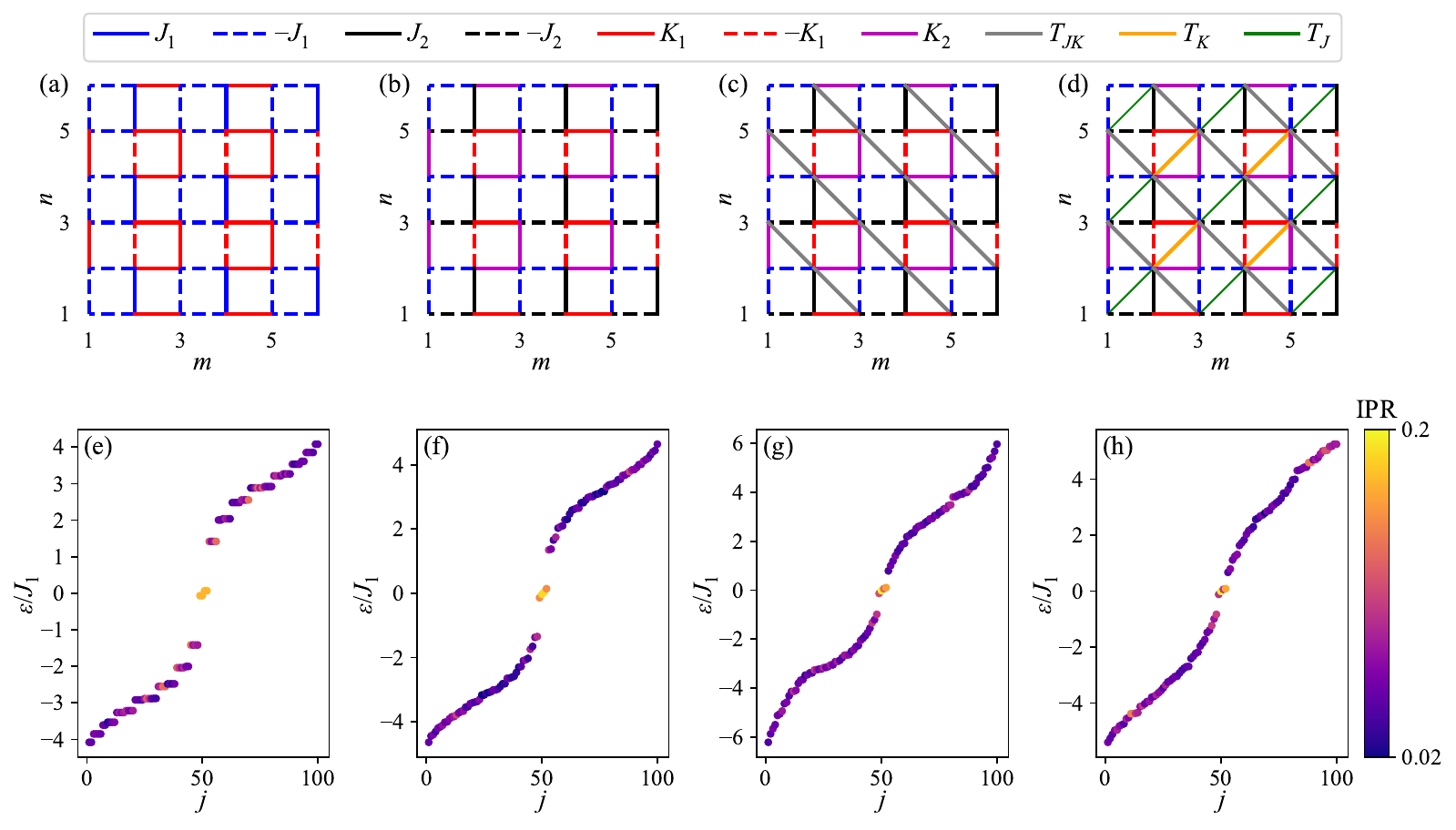}
  \caption{(a)-(d) Finite tight-binding models with couplings (a) $J$ and $K$, (b) $J_1$,$J_2$ and $K_1$,$K_2$, (c) the same as (b), but with diagonal couplings $T_{K}$ in the strong-coupling unit cell, (d) the same as (c), but will all diagonal couplings $T_{K}$, $T_{J}$, and $T_{JK}$. (e)-(h) Eigenvalue spectra for the models of (a)-(d) having the size of $10 \times 10$ sites with the following  parameters: $J_2=1.2J_1$, $K_1=2J_1$, $K_2=2.4J_1$, $T_{JK}=1.5J_1$, $T_{K}=J_1$, $T_{J}=0.5J_1$.}
  \label{fig:Extended_model}
\end{figure}

Next, we consider finite models with the same modifications. Compared to the original model of Fig.~\ref{fig:Extended_model}(a),(e), the spectrum of the model with couplings $J_{1}$,$J_{2}$ and $K_{1}$,$K_{2}$ shown in Fig.~\ref{fig:Extended_model}(b),(f) demonstrates an increased splitting of the corner state energies and is symmetric with respect to zero energy. Adding the diagonal couplings $T_{JK}$ between the unit cells breaks the symmetry of the spectrum and increases the splitting of the corner state energies, as seen in Fig.~\ref{fig:Extended_model}(c),(g). Such a diagonal intra-cell coupling is selected since the magnitude of the corresponding coupling should be higher compared to other diagonal couplings. Finally, introducing all the diagonal links $T_{K}$, $T_{J}$, and $T_{KJ}$ between the opposite sites with Type-1 resonators slightly changes the splitting of the corner state energies, but considerably increases the asymmetry of the spectrum and the difference between the upper and lower bands of the edge states, in a full agreement with the results of numerical simulations in Fig.~4 in the main text and Fig.~\ref{fig:Substrate_dispersion_and_spectra}. Thus, the observed differences between the numerical results and the model of Schulz and co-authors~\cite{2022_Schulz}, as well as a considerable splitting of corner state frequencies and the asymmetry of the edge state bands observed experimentally in Fig.~\ref{fig:Absorption} are related to the additional couplings between Type-1 resonators and associated chiral symmetry breaking.

\section{Fractional corner charges}
\label{sec:Topology}

The model shown in Fig.~\ref{fig:Extended_model}(a) features chiral symmetry manifested in the symmetry of the spectrum with respect to zero energy, Fig.~\ref{fig:Extended_model}(e), as well as projective reflection symmetries with respect to the $x$- and $y$-axes. These projective symmetries differ from ordinary ones by an additional transformation that flips the signs of the couplings in the unit cell by adding a phase factor to a wavefunction, owing to the presence of $\pi$-flux. These two reflection symmetries anticommute, which leads to the quantization of the bulk quadrupole moment~\cite{2017_Benalcazar_Science}. However, in the realized array of resonators there is an additional dimerization of couplings in the unit cell, which breaks all spatial symmetries except reflection with respect to one of the diagonals, but preserves chiral symmetry, Fig.~\ref{fig:Extended_model}(b),(f). Moreover, in the implemented array there are additional diagonal couplings between Type-1 SRRs, which break chiral symmetry, Fig.~\ref{fig:Extended_model}(c),(g). While the quadrupole moment remains quantized in models that either have chiral symmetry and a single reflection plane or lack chiral symmetry but have two non-commuting reflection symmetries, the model shown in Fig.~\ref{fig:Extended_model}(d),(h), which is the closest approximation for the experimental system, therefore cannot represent a quadrupole topological insulator.

\begin{figure}[t]
  \includegraphics[width=16cm]{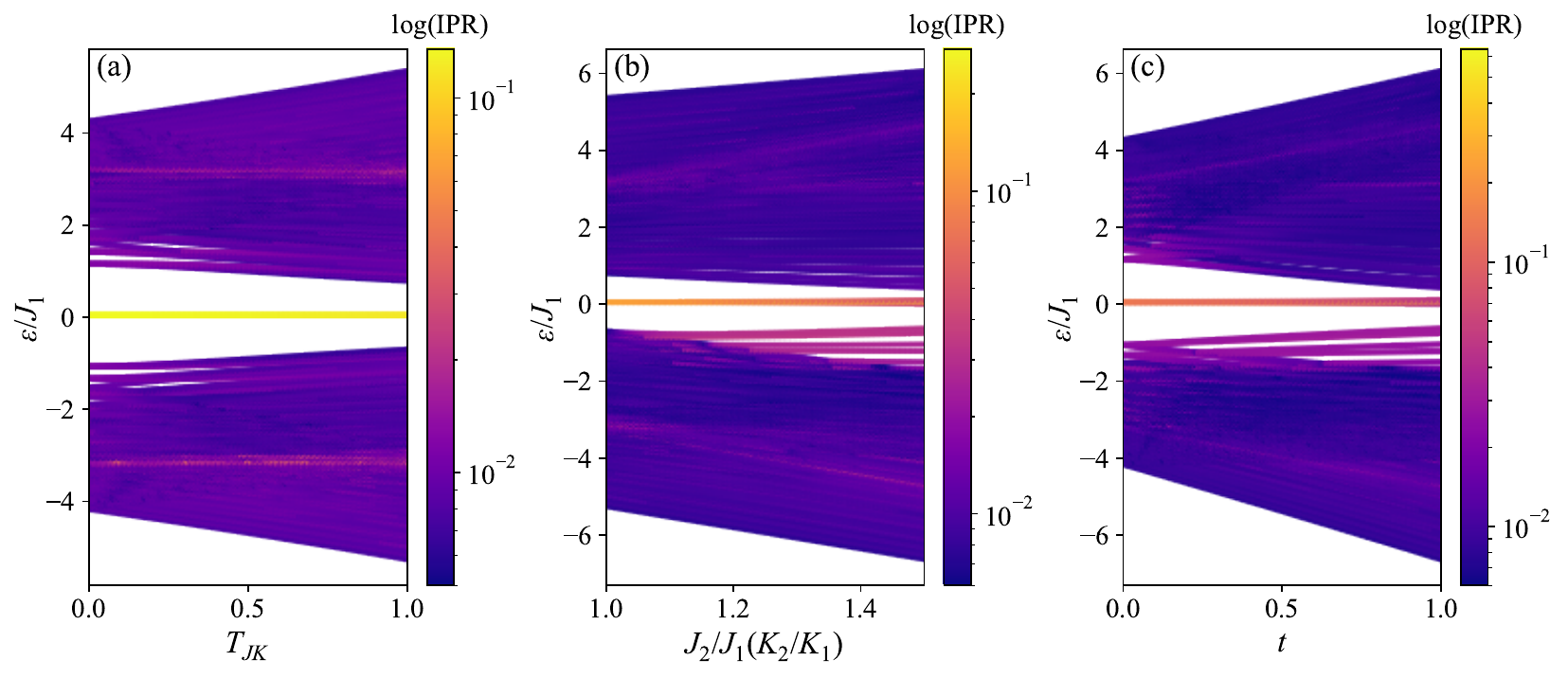}
  \caption{(a) Energy spectra of the model of Fig.~\ref{fig:Extended_model}(c) with $J_{2}/J_{1} = K_{2}/K_{1} = 1$ and $T_{JK}/J_{1}$ varied from $0$ to $1$. The color shows the values of inverse participation ratio $\rm {IPR}$ Eq.~(1). (b) Spectrum of the model with $T_{JK}/J_{1}=1$ and varying $J_{2}/J_{1} = K_{2}/K_{1}$ from $1$ to $1.5$. (c) Spectrum of the model under the simultaneous variation $J_{2}/J_{1} = K_{2}/K_{1} = 1 + t$ and $T_{JK}/J_{1} = t$ for values of $t$ from $0$ to $1$.}
  \label{fig:Variation}
\end{figure}

To study topological properties of the observed corner states, we consider an adiabatic variation of the parameters from the original Benalcazar-Bernevig-Hughes model of Fig.~\ref{fig:Extended_model}(a) to the model of Fig.~\ref{fig:Extended_model}(c), whose symmetry properties are the same as those of the model in Fig.~\ref{fig:Extended_model}(d). To do this, we vary the amplitude of the diagonal couplings $T_{JK}/J$ from $0$ to $1$, as shown in Fig.~\ref{fig:Variation}(a). It is seen that under such an adiabatic variation of the model parameters, the band gaps of the bulk and edge states remain open, and the system hosts highly localized in-gap corner states. When the additional dimerization is introduced and the couplings $J$ and $K$ are replaced by $J_{1,2}$ and $K_{1,2}$, chosen for simplicity in such a way that $J_{2}/J_{1} = K_{2}/K_{1}$, the band gaps remain open and the corner states are present, Fig.~\ref{fig:Variation}(b). Similarly, both parameters can be varied simultaneously, $J_{2}/J_{1} = K_{2}/K_{1} = 1 + t$ and $T_{JK}/J = t$, with $t$ varying from 0 to 1, Fig.~\ref{fig:Variation}(c). The persistence of the band gaps and the corner states indicates that they can still be linked to the initial topological system, despite the breaking of the symmetries that ensure the quantization of the bulk quadrupole moment, and thus may constitute a boundary-obstructed topological phase (BOTP)~\cite{2021_Khalaf}. Such states are characterized not by a symmetry indicator in the bulk, but by observable quantities at the boundary, such as fractional corner charges.

To demonstrate the presence of fractional corner charges in the considered model, we calculate the charge in each of the four quadrants of the model~\cite{2017_Benalcazar_PRB} as
\begin{equation}
   Q_{\text{corner}}^{(c)} = \sum_{m=1}^{N} \sum_{n=1}^{N} I_c(m,n)\,\big(\rho(m,n) - \frac{1}{2}\big) \pmod{1},
   \label{eq:Corner_charge}
\end{equation}
where $c \in \{(1,1), (1,N), (N,1), (N,N)\}$ is the corner index, $I_{c}$ is a function that equals $1$ in the quadrant between the corresponding site and the center of the model, and $0$ elsewhere, and $1\le m \le N, 1 \le n \le N$ are the site indices. The local probability density $\rho(m,n)$ at the site $(m,n)$ is defined
\begin{equation}
   \rho(m,n) = \sum_{j=1}^{N_{\text{sites}}/2} \left\lvert \beta_j(m,n) \right\rvert^2 .
\end{equation}
where $N_\text{sites} = N^2$ is the total number of model sites, and the wave functions $\beta_j = \beta(\varepsilon_j)$ are normalized by the condition
\begin{equation}
   \sum_{m=1}^{N}\sum_{n=1}^{N} \lvert \beta_j(m,n) \rvert^2 = 1,
\end{equation}
and the eigenenergies $\varepsilon_j$ are ordered by increasing value $\varepsilon_1 \leq \varepsilon_2 \leq ... \leq \varepsilon_{N_\text{sites}}$. The term $1/2$, which is subtracted from the local density in Eq.~\eqref{eq:Corner_charge}, represents the average bulk density corresponding to the $N_\text{sites}/2$ occupied states. For particular systems, instead of $N_{\text{sites}}/2$ one should include all corner states in the summation to obtain correct corner charge values.

Numerical calculations performed for the model of Fig.~\ref{fig:Extended_model}(a) with the parameters $N = 20$, $K/J = 2$ give the values of all four corner charges $Q(1,1)=Q(1,N)=Q(N,1)=Q(N,N)=0.500$, characteristic of a quadrupole insulator. This exact quantization is also preserved in the model with additional dimerization $J_{2}/J_{1}=K_{2}/K_{1}=1.2$ but without diagonal couplings, Fig.~\ref{fig:Extended_model}(b). At the same time, in the model of Fig.~\ref{fig:Extended_model}(c) with diagonal coupling $T_{JK}/J_1=1.5$, the values of the fractional charges are $Q(1,1)=0.505$, $Q(1,N)=0.439$, $Q(N,1)=0.552$, $Q(N,N)=0.505$. In the most complete model of Fig.~\ref{fig:Extended_model}(d) with the parameters $T_{JK}/J_1=1.5$, $T_J/J_1=0.5$, $T_K/J_1=1$, the calculated values of the corner charges are $Q(1,1)=0.515$, $Q(1,N)=0.955$, $Q(N,1)=0.016$, $Q(N,N)=0.515$. It is seen that the charges remain fractional, which, together with robustness of the band gaps and corner states, allows to classify the considered model as a higher-order BOTP.

Nevertheless, the values of the charges are not exactly equal to $0.500$, which indicates the absence of the quadrupole moment quantization and can be caused by the emergence of additional polarizations at the edges of the system. Importantly, the key symmetries are satisfied: $Q(1,1)=Q(N,N)$, which corresponds to the presence of a diagonal reflection symmetry in the model, and the sum of all four corner charges being integer. The obtained fractional charges are consistent with the dependence of the corner states on the type of boundary conditions of the model shown in Fig.~4(e) in the main text, as well as with the robustness of the corner states demonstrated experimentally.

To enhance the topological protection of the corner states, one can optimize the geometry and properties of the resonators so as to remove the additional dimerization and restore two reflection symmetries, or, alternatively, restore chiral symmetry by introducing screening elements that block the diagonal couplings between sites~\cite{2025_Huang}. Either route would enable the realization of a quadrupole topological phase.

\section{Absorption spectra in the experimental setup}
\label{sec:Absorption}

\begin{figure}[tbp]
  \includegraphics[width=12cm]{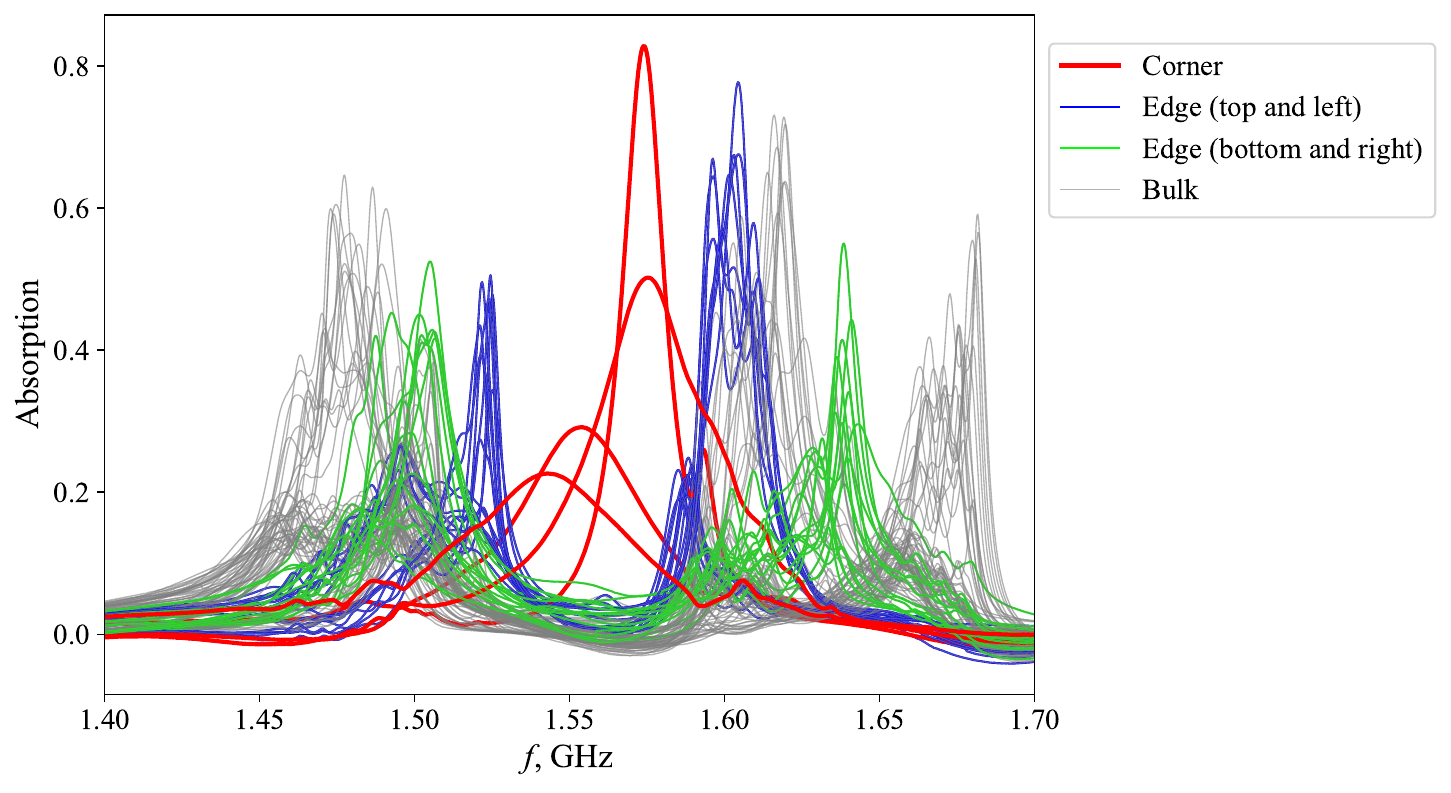}
  \caption{Experimentally measured absorption spectra for all sites in the array of $10 \times 10$ SRRs. The color indicates corner (the red solid lines), edge (the blue and green solid lines), and bulk (the gray solid lines) resonators.}
  \label{fig:Absorption}
\end{figure}

In this Section, we consider an array of $10 \times 10$ SRRs labeled by indices $(n, m)$. To estimate the absorption spectra for the system, we measure reflection coefficient $S_{11}$ using a loop antenna attached to the Planar S5085 vector network analyzer. The probe was oriented parallel to the given SRR and placed near the ring center. The resonances of Type-1 SRRs measured this way have a lower magnitude and a broader linewidth compared to resonances of Type-2 SRRs. The obtained absorption spectra, calculated as $1 - |S_{11}|^{2}$, are shown in Fig.~\ref{fig:Absorption}. Bulk resonators (the gray curves) feature resonant peaks at frequencies $1.45-1.50$~GHz (the bottom bulk band), as well as splitting resonances around frequencies $1.61$~GHz and $1.67$~GHz (the upper bulk bands), with a band gap at $1.50-1.59$~GHz separating these two regions. 

The resonators located at the bottom and right boundaries (the green curves) feature resonances at the frequencies around $f = 1.50$~GHz of the bottom and $f = 1.63$~GHz of the upper bulk regions. The resonators at the top and left boundaries possess resonances in the band gap at the frequencies around $f = 1.52$~GHz, as well as resonances overlapping with bulk states at the frequencies around $f = 1.60$~GHz. These results are consistent with the numerical simulations and measurements of the spatial distributions of reflection coefficient $S_{11}$ for the edge states discussed in the main text.

Finally, the spectra for the four corner resonators (the red curves) demonstrate resonances of Type-1 SRRs at $1.54$~GHz and $1.55$~GHz (corresponding to the corner state at the bottom left  and the top right corners shown in Fig.~5(e) in the main text) and a pair of resonances of Type-2 SRRs at $1.57$~GHz. Thus, the spectra clearly demonstrate the main features of the considered system: the presence of a bulk band gap, gapped edge states near the bottom bulk band, and corner states in the band gap that are spectrally separated from bulk and edge states.

The frequencies of the four corner states $f$, their relative offsets $\Delta f$, and their linewidths $\delta f$ are the following:
\begin{align*}
   \text{Corner (1, 1)}:& \quad f = 1.554~\text{GHz}, \quad \Delta f=80~\text{MHz}, \quad \delta f = 32~\text{MHz}\\
   \text{Corner (10, 1)}:& \quad f = 1.575~\text{GHz}, \quad \Delta f=130~\text{MHz}, \quad \delta f = 27~\text{MHz}\\
   \text{Corner (1, 10)}:& \quad f = 1.574~\text{GHz}, \quad \Delta f=120~\text{MHz}, \quad \delta f = 11~\text{MHz}\\
   \text{Corner (10, 10)}:& \quad f = 1.543~\text{GHz}, \quad \Delta f=190~\text{MHz}, \quad \delta f = 41~\text{MHz}
\end{align*}
The mean frequency of the four resonances is $f_{\text mean}=1.562$~GHz, which results in the calculated relative offset values $\Delta f$. It is seen that relative offsets are several times greater than linewidths, demonstrating that relative contributions of diagonal couplings between Type-1 resonators are dominant compared to losses.

\section{Fluctuations in parameters of individual resonators}
\label{sec:Fluctuations}

\begin{figure}[tbp]
  \includegraphics[width=15cm]{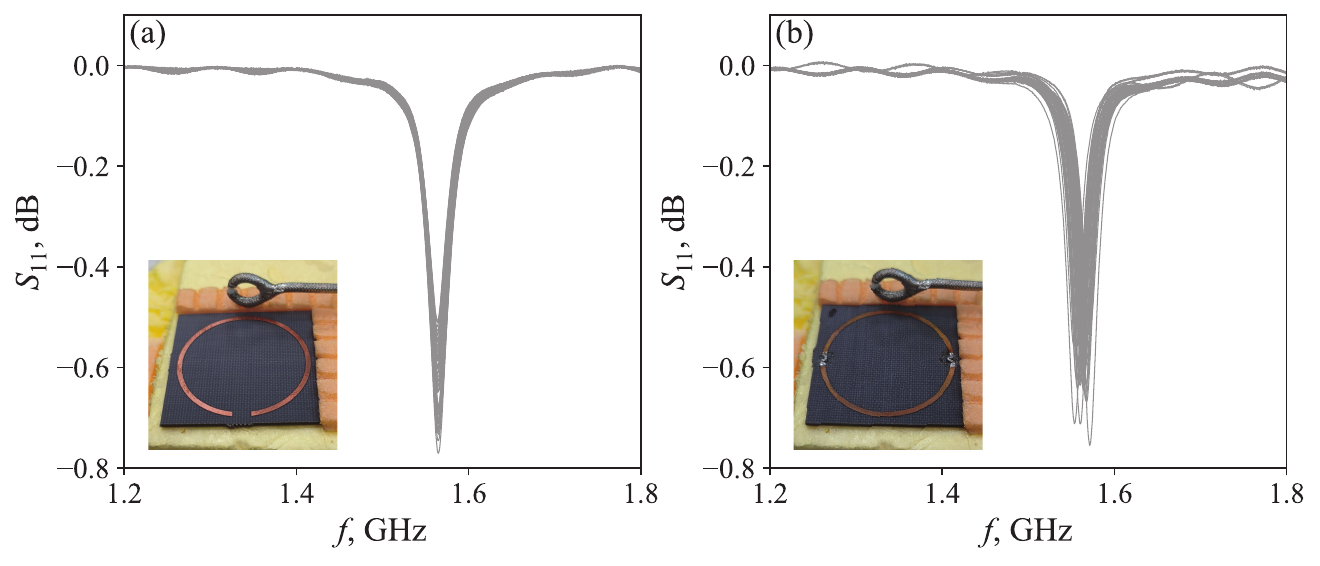}
  \caption{Experimentally measured reflection coefficient $S_{11}$ for (a) $50$ individual Type-1 SRRs and (b) $50$ individual Type-2 SRRs. The insets demonstrate examples of SRRs and the position of the magnetic probe.}
  \label{fig:SRR_S11}
\end{figure}

To estimate the diagonal disorder in the considered system, we measure the reflection coefficient $S_{11}$ for each of the individual SRRs with the help of a magnetic probe attached to the Planar~S5085 vector network analyzer. The position of the magnetic probe is shown in the insets of Fig.~\ref{fig:SRR_S11}(a),(b). The obtained spectra corresponding to $50$ Type-1 SRRs are shown in Fig.~\ref{fig:SRR_S11}(a) and feature the average resonant frequency $1.564$~GHz, while the frequencies of individual peaks range from $1.563$~GHz to $1.566$~GHz. At the same time, $S_{11}$ spectra of Type-2 SRRs possess considerably larger fluctuations in resonant frequencies compared to Type-1 SRRs. In particular, the positions of individual resonant peaks vary in the frequency range from $1.553$~GHz to $1.571$~GHz, with a mean frequency of $1.563$~GHz. Stronger fluctuations in the resonant frequencies of Type-2 SRRs are related to the $\pm 0.05$~pF tolerance in the capacitance of Murata GJM1555C1HR30WB01D and GJM1555C1HR60WB01D capacitors used in the experimental setup, as well as to manual soldering.

\section{Resilience towards boundary defects}
\label{sec:Zigzag}

\begin{figure}[b]
  \includegraphics[width=15cm]{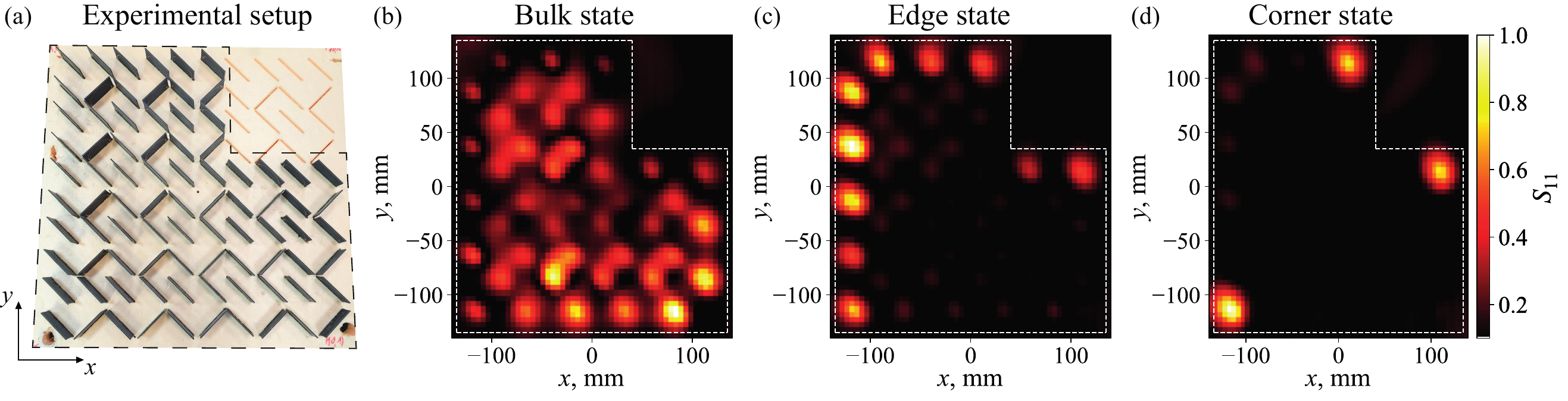}
  \caption{(a) Photo of the SRR array with a boundary defect. (b)-(d) Experimentally measured spatial distributions of reflection coefficient $S_{11}$ at the frequencies of (b) bulk ($f = 1.500$~GHz), (c) edge ($f=1.524$~GHz), and (d) corner ($f=1.550$~GHz) states.}
  \label{fig:Zigzag}
\end{figure}

In this Section, we experimentally demonstrate the robustness of edge and corner states to geometrical imperfections. To this end, we perform near-field measurements for the system with extracted region with $4 \times 4$ resonators at the top right corner of the structure of size $10 \times 10$ SRRs, Fig.~\ref{fig:Zigzag}(a), using the same reflection coefficient measurement method as discussed in the main text.

The considered system exhibits the same characteristic types of localization as the unperturbed experimental setup, including bulk ($f = 1.500$~GHz), edge ($f=1.524$~GHz), and corner ($f=1.550$~GHz) states, Fig.~\ref{fig:Zigzag}(b)-(d). In particular, the bulk state features delocalized distribution of reflection coefficient $S_{11}$, as shown in Fig.~\ref{fig:Zigzag}(b). At the frequencies of edge states, non-zero amplitudes of $S_{11}$ localize along the left and modified top boundaries of the SRR array, Fig.~\ref{fig:Zigzag}(c). However, the electromagnetic response along the modified top boundary is slightly weaker compared to that along the left edge. Finally, corner states demonstrate localization at three weak-link corners with Type-1 SRRs, with two of them achieved due to the introduced boundary modifications, Fig.~\ref{fig:Zigzag}(d).



%